\documentclass{ws-procs9x6-cpt22}
\begin{document}

\newcommand{\refeq}[1]{(\ref{#1})}
\def\etal {{\it et al.}}

\def\al{\alpha}
\def\be{\beta}
\def\ga{\gamma}
\def\de{\delta}
\def\ep{\epsilon}
\def\ve{\varepsilon}
\def\ze{\zeta}
\def\et{\eta}
\def\th{\theta}
\def\vt{\vartheta}
\def\io{\iota}
\def\ka{\kappa}
\def\la{\lambda}
\def\vpi{\varpi}
\def\rh{\rho}
\def\vr{\varrho}
\def\si{\sigma}
\def\vs{\varsigma}
\def\ta{\tau}
\def\up{\upsilon}
\def\ph{\phi}
\def\vp{\varphi}
\def\ch{\chi}
\def\ps{\psi}
\def\om{\omega}
\def\Ga{\Gamma}
\def\De{\Delta}
\def\Th{\Theta}
\def\La{\Lambda}
\def\Si{\Sigma}
\def\Up{\Upsilon}
\def\Ph{\Phi}
\def\Ps{\Psi}
\def\Om{\Omega}
\def\cA{{\cal A}}
\def\cB{{\cal B}}
\def\cC{{\cal C}}
\def\cE{{\cal E}}
\def\cl{{\mathcal L}}
\def\cL{{\mathcal L}}
\def\cO{{\cal O}}
\def\cP{{\cal P}}
\def\cR{{\cal R}}
\def\cV{{\cal V}}
\def\mn{{\mu\nu}}

\def\fr#1#2{{{#1} \over {#2}}}
\def\half{{\textstyle{1\over 2}}}
\def\quar{{\textstyle{1\over 4}}}
\def\frac#1#2{{\textstyle{{#1}\over {#2}}}}

\def\vev#1{\langle {#1}\rangle}
\def\bra#1{\langle{#1}|}
\def\ket#1{|{#1}\rangle}
\def\bracket#1#2{\langle{#1}|{#2}\rangle}
\def\expect#1{\langle{#1}\rangle}
\def\norm#1{\left\|{#1}\right\|}
\def\abs#1{\left|{#1}\right|}

\def\lsim{\mathrel{\rlap{\lower4pt\hbox{\hskip1pt$\sim$}}
    \raise1pt\hbox{$<$}}}
\def\gsim{\mathrel{\rlap{\lower4pt\hbox{\hskip1pt$\sim$}}
    \raise1pt\hbox{$>$}}}
\def\sqr#1#2{{\vcenter{\vbox{\hrule height.#2pt
         \hbox{\vrule width.#2pt height#1pt \kern#1pt
         \vrule width.#2pt}
         \hrule height.#2pt}}}}
\def\square{\mathchoice\sqr66\sqr66\sqr{2.1}3\sqr{1.5}3}

\def\prt{\partial}

\def\etal{{\it et al.}}

\def\pt#1{\phantom{#1}}
\def\ni{\noindent}
\def\ol#1{\overline{#1}}

\def\ss{$s^{\mu\nu}$}
\def\tt{$t^{\ka\la\mu\nu}$}
\def\uu{$u$}

\def\sss{s^{\mu\nu}}
\def\ttt{t^{\ka\la\mu\nu}}

\def\sb{\overline{s}}
\def\tb{\overline{t}}
\def\ub{\overline{u}}

\def\stw{\tilde{s}}
\def\ttw{\tilde{t}}
\def\utw{\tilde{u}}
\def\Btw{\tilde{B}}

\def\hsy{h_{\mu\nu}}  
\def\nsy{\et_{\mu\nu}}

\def\nsc#1#2#3{\om_{#1}^{{\pt{#1}}#2#3}}
\def\lsc#1#2#3{\om_{#1#2#3}}
\def\usc#1#2#3{\om^{#1#2#3}}
\def\lulsc#1#2#3{\om_{#1\pt{#2}#3}^{{\pt{#1}}#2}}

\def\tor#1#2#3{T^{#1}_{{\pt{#1}}#2#3}}

\def\vb#1#2{e_{#1}^{{\pt{#1}}#2}}
\def\ivb#1#2{e^{#1}_{{\pt{#1}}#2}}
\def\uvb#1#2{e^{#1#2}}
\def\lvb#1#2{e_{#1#2}}

\def\barvb#1#2{\bar e_{#1}^{{\pt{#1}}#2}}
\def\barivb#1#2{\bar e^{#1}_{{\pt{#1}}#2}}
\def\baruvb#1#2{\bar e^{#1#2}}
\def\barlvb#1#2{\bar e_{#1#2}}

\newcommand{\beq}{\begin{equation}}
\newcommand{\eeq}{\end{equation}}
\newcommand{\bea}{\begin{eqnarray}}
\newcommand{\eea}{\end{eqnarray}}
\newcommand{\bit}{\begin{itemize}}
\newcommand{\eit}{\end{itemize}}
\newcommand{\rf}[1]{(\ref{#1})}

\title{Explicit Diffeomorphism Breaking, No-go Conditions, \\
and the Gravity Sector of the SME}

\author{R.\ Bluhm$^1$}

\address{$^1$Department of Physics and Astronomy, Colby College,\\
Waterville ME 04901, USA}


\begin{abstract}
A brief recap is given of issues that can occur when fixed nondynamical backgrounds
that explicitly break diffeomorphism invariance are included in gravity theories.
Implications for the Standard-Model Extension (SME) are summarized,
including recent results showing how the SME can be modified to accommodate such
fixed nondynamical backgrounds.
\end{abstract}

\bodymatter

\section{Introduction}

When the gravity sector of the SME was first developed in 2004,
it was found that with explicit breaking of diffeomorphisms and local
Lorentz invariance inconsistencies between geometric identities,
dynamics, and covariant energy-momentum conservation can occur,
while with spontaneous breaking these inconsistencies are evaded.\cite{akgrav04}
The occurrence of such inconsistencies in the case of explicit breaking
was referred to as a no-go condition.

Explicit breaking occurs when a nondynamical background field is included
in the theory, while with spontaneous breaking all the fields are dynamical,
and consistency with geometric identities such as the Bianchi identities is assured.
As a result of the no-go condition in the case of explicit breaking,
the original development of the gravity sector of
the SME assumed that the symmetry breaking was spontaneous.

In the subsequent years after the gravity sector of the SME was introduced, 
many features of spontaneous spacetime symmetry
breaking, such as the appearance of Nambu-Goldstone (NG) 
and massive Higgs-like excitations, were explored.\cite{rbak}
These features then played a key role in developing the post-Newtonian limit
of the SME as well as in finding systematic procedures for including matter-gravity couplings,
which has led to many experimental bounds being placed on 
gravitational interactions that break spacetime symmetry.\cite{akqbjt}

At the same time, however, gravity theories with explicit breaking of spacetime
symmetries were being developed, 
such as massive gravity and Horava gravity.
This led to a reexamination of explicit breaking and the question of whether
the SME can be used to investigate theories with nondynamical backgrounds.\cite{rb}
Most recently, in 2021, an analysis of how to extend the SME to include nondynamical backgrounds 
was made and investigations of the phenomenological implications of this have commenced.\cite{akzl}

For simplicity in this brief review, only explicit breaking of diffeomorphisms
is considered.  
Results concerning the explicit breaking of local Lorentz symmetry
can be found in the references.

\section{SME with Explicit Breaking}

A gravity theory with a fixed background tensor,
which can be represented generically as $\bar \ta^\mu_{\pt{\mu}\nu}$,
has an action of the form
\beq
S_{\rm SME} = {\Huge \int} d^4x \sqrt{-g} \, \left[   \fr 1 {2} R +
{\cal L} (g_{\mu\nu}, \bar \ta^\mu_{\pt{\mu}\nu}, \dots) \right] ,
\label{S}
\eeq
with $8 \pi G = 1$.  
In the case of the SME,
the background $\bar \ta^\mu_{\pt{\mu}\nu}$ could represent an SME coefficient.

With spontaneous diffeomorphism breaking, the background $\bar \ta^\mu_{\pt{\mu}\nu}$
is understood to be a vacuum expectation value $\bar \ta^\mu_{\pt{\al}\nu} = \vev{\ta^\mu_{\pt{\al}\nu}}$
of a fully dynamical field $\ta^\mu_{\pt{\mu}\nu}$.
In this case, $\bar \ta^\mu_{\pt{\mu}\nu}$ obeys the vacuum equations of motion
along with a vacuum solution for the metric.  
When the NG excitations about the vacuum solution are included, 
the action $S$ is diffeomorphism invariant.
The symmetry becomes hidden (not broken) with spontaneous breaking.

In contrast, with explicit diffeomorphism breaking, the background $\bar \ta^\mu_{\pt{\mu}\nu}$
is a fixed nondynamical background that is inserted in the action.  
It has no excitations and does not have equations of motion.  
When diffeomorphisms are performed,
the metric transforms dynamically with changes given by the Lie derivative.
However, $\bar \ta^\mu_{\pt{\mu}\nu}$ remains fixed under diffeomorphims
and does not transform.
As a result the action $S$ is not invariant under diffeomorphisms.

Despite the fact that $(\de S)_{\rm diffs} \ne 0 $ under idiffeomorphisms,
the action must still be invariant under general coordinate transformations
in order to maintain observer independence.
Thus, $(\de S)_{\rm GCTs} = 0$ even when $(\de S)_{\rm diffs} \ne 0 $.
By considering infinitesimal general coordinate transformations,
where both the metric and $\bar \ta^\mu_{\pt{\mu}\nu}$ transform passively
with changes given by their Lie derivatives,
four off-shell Noether identities can be obtained.
By combining these with the contracted Bianchi identities
the four Noether identities can be written symbolically as 
\beq
D_\mu T^{\mu\nu} = ( \cdots \fr {\de S} {\de \bar \ta^{\al}_{\pt{\mu}\be} } 
\cdots D_\al \fr {\de S} {\de \bar \ta^{\al}_{\pt{\mu}\be} } \cdots )^\nu  ,
\label{DT}
\eeq
where $v=0,1,2,3$ for the four identities, 
and $\fr {\de S} {\de \bar \ta^{\mu}_{\pt{\mu}\nu}}$ represents the Euler-Lagrange expression
for the background $\bar \ta^\mu_{\pt{\mu}\nu}$.
Evidently, consistency with the Einstein equations and covariant energy-momentum
conservation requires that the right-hand side in \rf{DT} must vanish on shell.

With spontaneous breaking, $\fr {\de S} {\de \bar \ta^{\mu}_{\pt{\mu}\nu}} = 0$  because 
this represents the vacuum equation of motion.
However, with explicit breaking there is no reason for $\fr {\de S} {\de \bar \ta^{\mu}_{\pt{\mu}\nu}}$
to vanish, since $\bar \ta^\mu_{\pt{\mu}\nu}$ does not obey equations of motion.
Hence, an inconsistency can arise such that $D_\mu T^{\mu\nu} \ne 0$,
which is the no-go result.

It is nonetheless possible to evade the no-go result.\cite{rbak}
This is because in a theory with explicit breaking
there are four extra degrees of freedom due to
the loss of local diffeomorphism invariance.
As long as these four additional degrees of freedom are not suppressed or decouple,
they can take values that can make the right-hand side of \rf{DT} vanish.
In fact, a St\"uckelberg approach can be used to reveal the form of the four
extra degrees of freedom that occur with explicit breaking.
The result is that the St\"uckelberg excitations have the same
form as the NG modes in a corresponding theory with spontaneous breaking.
Hence, the St\"uckelberg approach introduces precisely the four NG modes that
are minimally required to maintain the consistency of the theory and to evade the
no-go result.

The implications of using the SME with explicit breaking and where the SME
coefficients can be interpreted as nondynamical backgrounds can then be examined
for both the pure-gravity sector in a linearized post-Newtonian limit and 
also for matter-gravity couplings.  
However, since the St\"uckelberg excitations have the form of
NG modes and the linearized curvature tensor $R^\ka_{\pt{\ka}\la\mu\nu}$
is gauge invariant, the NG modes decouple at leading order.
Aa a result, there is no useful linearized post-Newtonian limit.
In contrast, when matter-gravity couplings are included,
the background SME coefficients in general have sufficient
couplings to matter terms in the action that are 
not gauge invariant at leading order.
Hence, the NG modes do not decouple,
and the same procedures that were carried out in the 
case of spontaneous breaking
can again be applied when the symmetry breaking is explicit.
It is therefore possible to use the SME to investigate matter-gravity
couplings in theories with explicit breaking.

In fact, a comprehensive extension of the SME has recently been
constructed for gravity theories with nondynamical backgrounds and
explicit breaking.\cite{akzl}
In making this construction, 
care must be taken to treat backgrounds with different types
of indices, such as $\bar \ta^\mu_{\pt{\mu}\nu}$, $\bar \ta_\mu^{\pt{\mu}\nu}$,
$\bar \ta^{\mu\nu}$, and $\bar \ta_{\mu\nu}$
as different theories,
since raising and lowering indices on nondynamical backgrounds cannot
be performed using the dynamical metric.  
Instead, each of these backgrounds can couple to matter differently and can
cause different symmetry-breaking signatures. 
The no-go conditions for them differ as well.
Likewise,
indices in a local Lorentz frame, $\bar \ta^a_{\pt{a}b}$, give rise to
different effects,
since the dynamical vierbein cannot be used to obtain these
components from $\bar \ta^\mu_{\pt{\mu}\nu}$.
However, in this case indices can be raised or lowered in the
local frame using the Minkowski metric $\et_{ab}$.

It is argued that perturbatively the no-go conditions cannot in general
be satisfied in the new extension of the SME.\cite{akzl}
The reasoning is that the required extra excitations needed to satisfy
the no-go conditions are either missing in many scenarios, 
or since gravity is so weak the backgrounds would need to have 
unnatural values, such as that they must be constant throughout
the gravitational field.
As an extension of this argument, 
searches for evidence of fixed backgrounds where the
no-go results are not satisfied can be viewed effectively as 
searches for beyond-Riemann geometry.

\section{Conclusion}

In summary, the SME has been reinterpreted and expanded in such a way 
that allows applications to gravity theories with nondynamical backgrounds
and explicit diffeomorphism breaking.  
Its overall consistency is limited in perturbative treatments due to the inability
to satisfy the no-go conditions when the St\"uckelberg NG modes decouple.  
Nonetheless, the new extension can be applied as a phenomenological
framework in experiments searching for effects of fixed backgrounds
in modified gravity theories.

\end{document}